\documentclass[showpacs,aps,prl,twocolumn,superscriptaddress,10pt]{revtex4-2}
\usepackage{graphicx} % Include figure files
\usepackage{bm}% bold math
\usepackage{color}

\usepackage{amsmath}
\usepackage{amssymb}
\usepackage{enumerate}
\usepackage{xspace}
\usepackage{mathrsfs}
\usepackage{mathptmx}
\usepackage[utf8]{inputenc}

\newcommand{\sS}{\ensuremath{\mathbf{S}}\xspace}

\newcommand{\smmu}{\ensuremath{\mu_{\mathrm{s}}}\xspace}

\newcommand{\smB}{\ensuremath{\mathbf{B}}\xspace}

\newcommand{\sms}{\ensuremath{\mathbf{S}}\xspace}
\newcommand{\Jij}{\ensuremath{J_{ij}}\xspace}
\newcommand{\Jnn}{\ensuremath{J_{ij}^{\mathrm{nn}}}\xspace}
\newcommand{\Jnnn}{\ensuremath{J_{ij}^{\mathrm{nnn}}}\xspace}

\newcommand{\IrMn}{\ensuremath{\text{IrMn}_3}\xspace}
\newcommand{\Neel}{N\'eel\xspace}

\newcommand{\etal}{\textit{et al}\xspace}

\begin{document}

\title{The atomistic origin of the athermal training effect in granular IrMn/CoFe bilayers}

\author{Sarah Jenkins}
\email{sarah.jenkins@york.ac.uk}
\affiliation{Department of Physics, University of York, York, YO10 5DD, UK}
\author{Roy.~W.~Chantrell}
\affiliation{Department of Physics, University of York, York, YO10 5DD, UK}
\author{Richard.~F.~L. Evans}
\email{richard.evans@york.ac.uk}
\affiliation{Department of Physics, University of York, York, YO10 5DD, UK}

\begin{abstract}
Anti-ferromagnetic materials have the possibility to offer ultra fast, high data density spintronic devices. A significant challenge is the reliable detection of the state of the antiferromagnet, which can be achieved using exchange bias. Here we develop an atomistic spin model of the athermal training effect, a well known phenomenon in exchange biased systems where the bias is significantly reduced after the first hysteresis cycle. We find that the setting process in granular thin films relies on the presence of interfacial mixing between the ferromagnetic and antiferromagnetic layers. We systematically investigate the effect of the intermixing and find that the exchange bias, switching field and coercivity all increase with increased intermixing. The interfacial spin state is highly frustrated leading to a systematic decrease in interfacial ordering of the ferromagnet. This metastable spin structure of initially irreversible spins leads to a large effective exchange coupling and thus large increase in the switching field. After the first hysteresis cycle these metastable spins drop into a reversible ground state that is repeatable for all subsequent hysteresis cycles, demonstrating that the effect is truly athermal. Our simulations provide new insights into the role of interface mixing and the importance of metastable spin structures in exchange biased systems which could help with the design an optimisation of antiferromagnetic spintronic devices.
\end{abstract}

%\pacs{75.10.Hk,75.20.-g,75.50.Ss,75.60.Jk,75.78.Jp}
\maketitle

\section{Introduction}

Anti-ferromagnetic spintronic devices have the potential to greatly outperform conventional ferromagnetic devices, due to the possibility for high data density storage and ultra fast dynamics~\cite{Jungwirth2016AntiferromagneticSpintronics}. These anti-ferromagnetic spintronic devices use the anti-ferromagnet (AFM) to store and transmit information. High data densities are possible due to the lack of stray fields in the AFM, however, without these fields it is very difficult to control and detect the magnetisation. The AFM magnetisation can be stimulated and detected electrically~\cite{Godinho2018ElectricallyAntiferromagnet,Zelezny2018SpinDevices,Asa2020ElectricalEffect}, but the read-out signals are still small at room temperature. A promising way of detecting and controlling the magnetisation is through the exchange bias effect, which occurs when the AFM is coupled to a ferromagnet(FM). The AFM causes the hysteresis loop of the FM to be shifted away from the zero field point with a preferred unidirectional orientation due to exchange anisotropy~\cite{Meiklejohn1957NewAnisotropy}. This effect has been used to demonstrate 180 degree switching of an antiferromagnet~\cite{LinNatMat2019} done using spin orbit torques. To gain full control of the switching the exchange bias needs to be optimised and tuned. 

One problem in the optimisation of the exchange bias effect is the training effect. The training effect causes a large drop in the measured exchange bias after the first hysteresis loop~\cite{Qiu2008RotationBilayers}, which continues with successive hysteresis cycles. Fernandez-Outen \etal~\cite{Fernandez-Outon2004ThermalSystems} postulated that the training effect could be split into two types of training: thermal training and athermal training. Thermal training is due to thermally activated depinning of the uncompensated AFM spins, usually causing a small change in the exchange bias and coercivity between every hysteresis loop~\cite{Kaeswurm2011TheFilms}. Athermal training is characterised by an abrupt decrease of coercivity and exchange bias between the first and second measured hysteresis loops. 

Thermal training is due to well understood thermal instabilities in the AFM~\cite{Qi2019InfluenceMultilayers}. The origin of athermal training however is still a widely disputed problem due to the difficulty in experimentally probing the rearrangement of AFM spins at the interface. It has been proposed to be due to the degree of order of the AFM at the interface~\cite{Biternas2009StudyModel} or changes in the configuration of the antiferromagnet during the hysteresis cycle~\cite{HoffmannPRL2004,DeClercqJPhysD2016,BremsPRL2007,MoritzPRB2016,ZhouSciRep2015}. In IrMn systems the magnetic anisotropy is extremely large~\cite{Szunyogh2009GiantPrinciples,Jenkins2019MagneticIrMn3} and so re-orientation of the bulk antiferromagnetic spins during the hysteresis cycle is not possible, and so typical micromagnetic approaches used to model exchange bias in Co/CoO do not apply.

%The initial cooling produces an AFM spin structure which may be a meta stable state. The athermal training effect can then be considered to be the rearrangement of the spin structure towards the minimum energy state during the first hysteresis loop. Biternas \etal predicted that training occurs due to meta-stable spins, "These spins are in a meta-stable state created during the setting process. During the first reversal, they reverse only once and they are pinned to a new easy direction"~\cite{Biternas2009StudyModel}. This can be seen by the change in shape of the hysteresis loops in Fig. \ref{fig:train}. The first branch of the first hysteresis loop is very square whereas the first branch of the second hysteresis loop is more rounded. The change in shape of these first branches suggests that the AFM layer is in a higher energy meta stable state after setting and transitions to a lower energy state during the hysteresis loop~\cite{Biternas2009StudyModel}. The return branch of both the first and second hysteresis loops are very similar suggesting that the transition actually occurs either after or during the first branch but before the second branch.

Recently, an alternative model of exchange bias for $\gamma$-\IrMn / CoFe bilayers has been proposed by Jenkins \etal~\cite{JenkinsEB2020}, including a realistic $3Q$ tetrahedral spin structure. The model gave accurate values for the exchange bias loop shift and the increase in coercivity due to the coupling with the ferromagnet. They found the exchange bias originates from the natural structural disorder in IrMn, creating a small statistical imbalance in the number of interfacial spins. Their model explains the origin of exchange bias without the need for AFM domains or impurities. In further work~\cite{JenkinsGranular2020} it was found that the exchange bias in granular IrMn/CoFe systems is of a similar magnitude, but that magnetic impurities led to a natural magnetic texture and dispersion of set directions for the film. 

Here we consider the athermal training effect in IrMn/CoFe bilayers by performing large scale atomistic simulations of the exchange bias. We simulate the field cooling setting procedure to initialise the exchange bias, subsequent athermal relaxation and then a slow hysteresis calculation with critical damping to compute the exchange bias, switching field and coercivity. We systematically investigate the role of the degree of interface mixing on the exchange biased properties and study the resulting microscopic spin structures to determine the origin of the athermal training effect in IrMn/CoFe systems.

\section{Method}
The simulations were performed using the \textsc{vampire} software package~\cite{Evans2014}. The simulations used an atomistic spin model with the energetics of the system being described by the spin Hamiltonian:
\begin{eqnarray}
\mathscr{H} &=& -\sum_{i<j} \Jij \sS_i \cdot \sS_j - \frac{k_N}{2} \sum_{i \neq j}^z (\mathbf{S}_i \cdot  \mathbf{e}_{ij})^2 \nonumber \\
&& - \sum_i k_{\mathrm{u}} (\sS \cdot \mathbf{e}_z)^2 - \sum_i \smmu \sms_i \cdot \smB,
\label{eq:hamiltonian}
\end{eqnarray}
where $\sS_i$ is the spin direction on site $i$, $k_N = -4.22 \times 10^{-22}$ is the \Neel pair anisotropy constant and $\mathbf{e}_{ij}$ is a unit vector from site $i$ to site $j$, $z$ is the number of nearest neighbours, \Jij is the exchange interaction and $\smB$ the strength of the external magnetic field. The effective exchange interactions in the 5 nm thick IrMn layer were limited to nearest ($\Jnn = -6.4 \times 10^{-21}$ J/link) and next nearest ($\Jnnn = 5.1 \times 10^{-21}$ J/link) neighbours~\cite{Jenkins2018EnhancedFilms,Jenkins2019MagneticIrMn3}. 
The 3 nm CoFe layer is simulated with a nearest neighbour approximation and a weak easy-plane anisotropy $k_{\mathrm{u}}$ to simulate the effects of the demagnetising field of a thin film. The exchange coupling across the FM/AFM interface is set at 1/5th of the bulk exchange values as calculated by \textit{ab-initio} methods~\cite{Szunyogh2011AtomisticInterface}. Spin Dynamics simulations were done solving the stochastic Landau-Lifshitz-Gilbert equation with a Heun numerical scheme~\cite{Ellis2015TheModels}. Our model naturally reproduces the low temperature ground state spin structures where the ordered alloy forms a triangular (T1) spin structure with an angle of 120 degrees between adjacent spins and the disordered alloy forms a tetrahedral (3Q) spin structure with 109.5 degrees between spins \cite{Jenkins2018EnhancedFilms} in agreement with previous neutron scattering experiments~\cite{Tomeno1999MagneticMn3Ir, Kohn2013TheExchange-bias.} and theoretical calculations~\cite{Szunyogh2011AtomisticInterface,Sakuma2003First-principlesAlloys,Hemmati2012MonteLattice}. The simulations also reproduce the \Neel ordering temperature of  730K for the disordered $\gamma$ phase~\cite{Yamaoka1974AntiferromagnetismAlloys}. The simulations were run in parallel on 400 cores to enable 20 ns timescale hysteresis loops, ensuring a converged coercivity and value for the exchange bias (in the limit of critical damping $\alpha = 1$).

\section{Results}
The system generated is described by Jenkins \etal~\cite{JenkinsGranular2020}. The simulated structure was 50 nm $\times$ 50 nm $\times$ 8 nm thin film, where the grain distribution had a median grain size of 5.5 nm and a lognormal standard deviation of 0.37. Experimentally, for exchange bias to occur the system needs to be heated and then cooled in the presence of a high-strength magnetic field~\cite{OGrady2010AFilms}. During this step the net direction of the uncompensated spins at the FM/AFM interface align with the field. It was discovered by Jenkins~\etal~\cite{JenkinsGranular2020} that, due to the finite timescales of the simulation, the simulated system has an equal probability of setting into any of the 8 possible ground states and therefore in a multigranular system each grain is set in a different direction and not along the setting field. We therefore initially use the setting procedure described by Jenkins~\etal~\cite{JenkinsGranular2020} to set the exchange bias in the system.

\begin{figure}[!tb]
\centering
\includegraphics[width=0.5\textwidth]{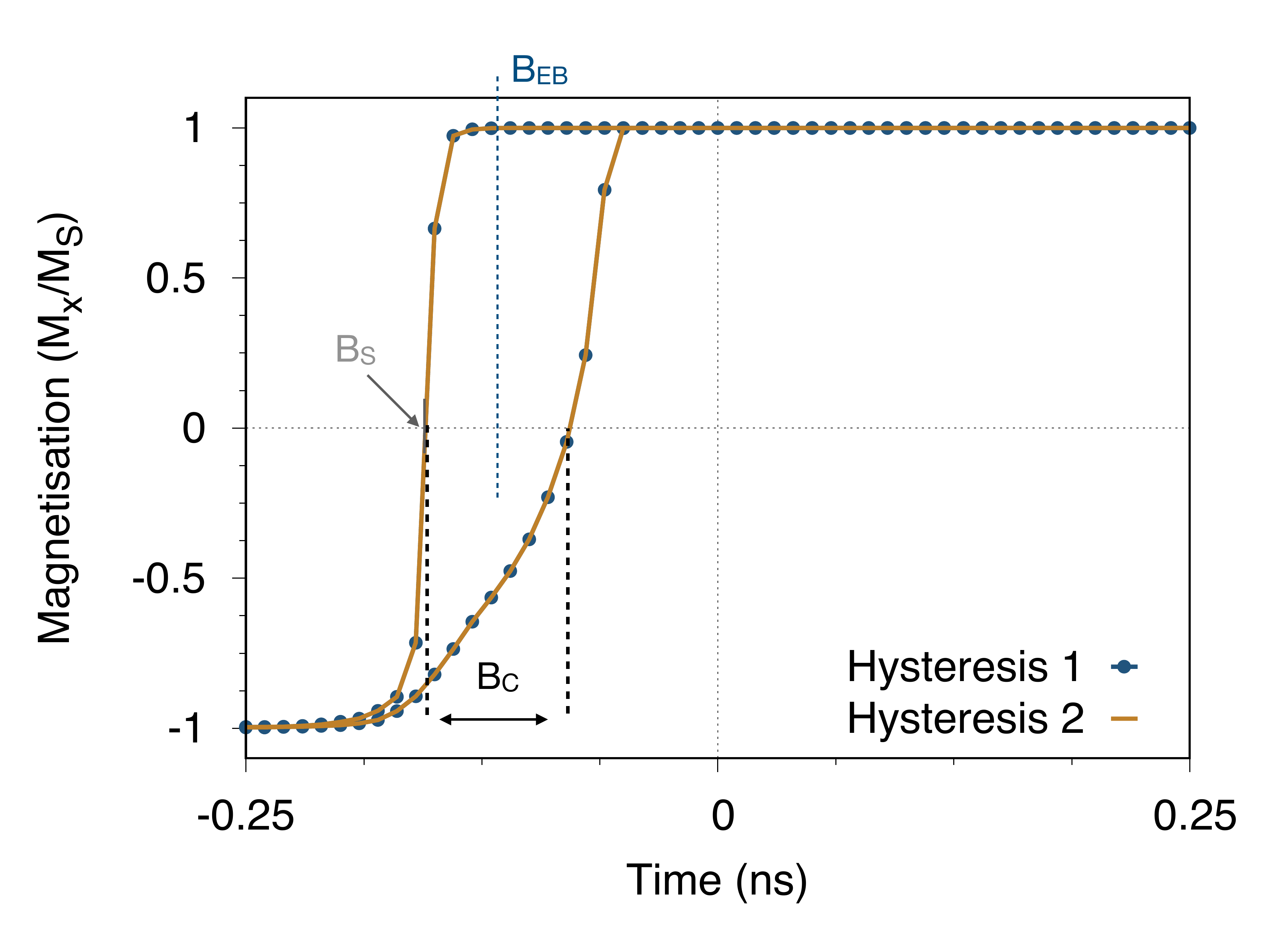}
\caption{\textbf{First and second simulated hysteresis loops at $T= 0$ K for a multigranular IrMn\textbackslash CoFe system}. The system does not exhibit the training effect with both hysteresis loops being identical with an exchange bias $\mu_0 H_{\mathrm{ex}} = 0.12$ T and coercivity $\mu_0 H_{\mathrm{c}} = 0.08$ T.}
\label{fig:train_sg_flat}
\end{figure}

We wish to propose that the athermal training effect in IrMn systems occurs due to disorder at the interface. To prove this theory, initially the system is set to have no interface disorder. The calculated first and second simulated hysteresis loops are shown in Fig.~\ref{fig:train_sg_flat}. Both hysteresis loops have an exchange bias of 0.12 T and a coercivity of 0.08 T. There is also no change in the shape of the loop between the first and second measured hysteresis loops. The lack of training is not surprising as Kaeswurm \etal~\cite{Kaeswurm2011TheFilms} predicted the training effect to be due to disorder of the Mn at the interface and the interface simulated here is completely atomically flat. To test her theory we will add some disorder to the interface. The disorder will be added by mixing the atoms at the interfaces. The first step is to create a bilayer system with a mixed interface~\cite{Jenkins2018EnhancedFilmsb}. The procedure for creating a mixed interface bilayer system is outlined below. Experimentally, the interface mixing in IrMn/CoFe systems has been measured to be in the region of 0.1nm - 1nm in width~\cite{Kanak2008InfluenceMultilayers,Qi2019InfluenceMultilayers}. To create a disordered interface, the material type (CoFe or Mn) was randomly swapped (CoFe to Mn and Mn to CoFe) around the interface. The swapping was generated using a probability distribution defined by:
%--
\begin{equation}
P(z) = 1 -\frac{1}{2} \tanh{\left(\frac{\pi(z-z_0)}{w}\right)} ,
\label{eq:mix}
\end{equation}
%--
\noindent where $P(z)$ is the probability of finding an atom of a particular type at height $z$, $z_0$ is the interface height and $w$ is the width of the tanh function, corresponding to the width of the interface mixing in nanometers. Every atom in the IrMn layer has a probability (P) of being changed to a CoFe atom depending on its height ($z$) above the interface. The mixing can also occur the other way around mixing the CoFe into the Mn or both types of mixing can occur simultaneously. Iridium has a very high atomic weight in comparison to CoFe therefore it is expected that the CoFe will not be able to penetrate into the IrMn layer, but will instead the Mn penetrate into the CoFe. The choice of diffusion type matches previous experimental measurements of the diffusion~\cite{Lee2002InterdiffusionMultilayer}. 

\begin{figure}[!tb]
\centering
\includegraphics[width=0.5\textwidth]{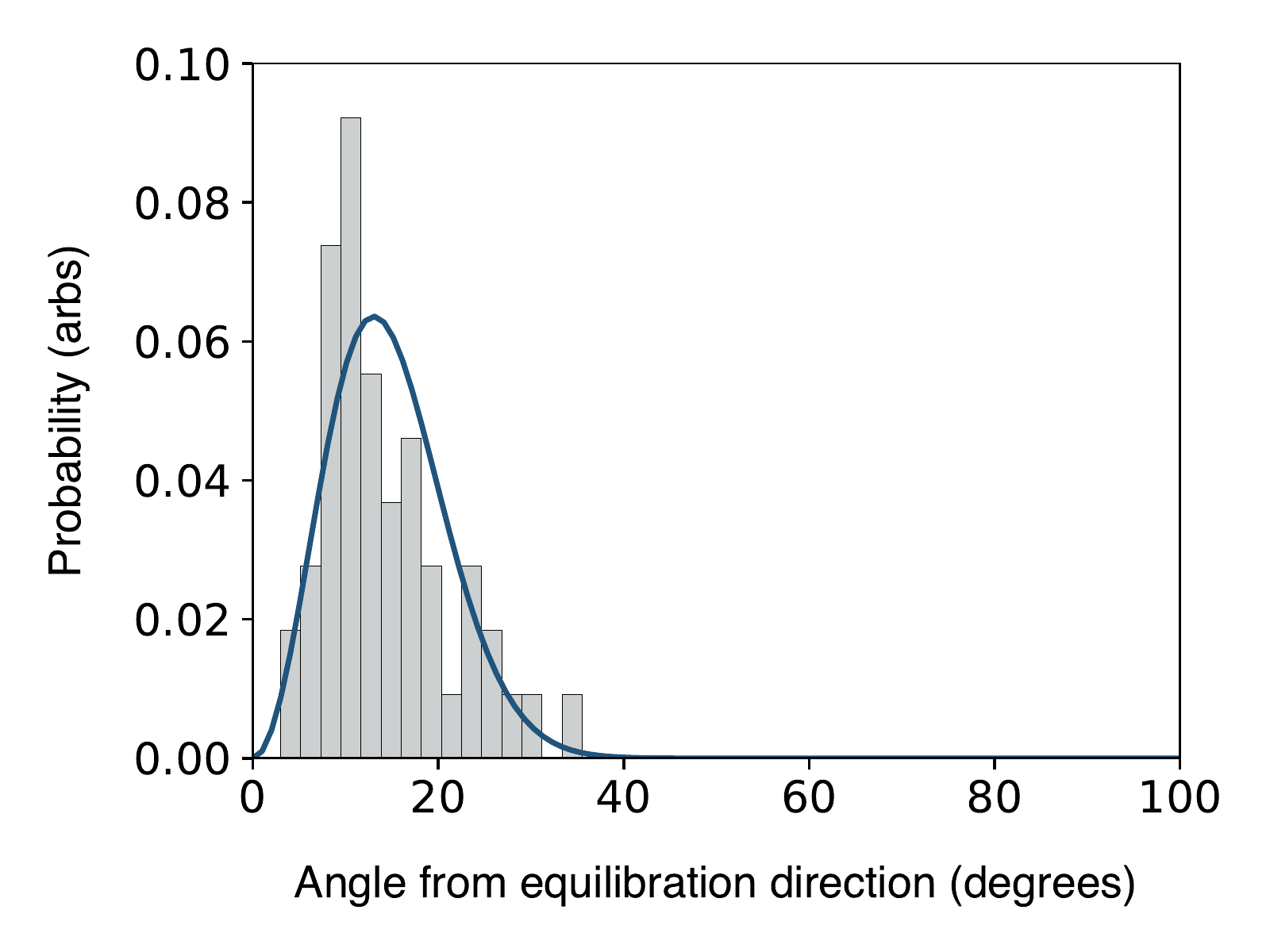}
\caption[Rotation from the setting field direction after the equilibration simulation]{\textbf{Rotation of the CoFe from the setting field direction after the equilibration simulation.} The CoFe has rotated to a maximum of 35 degrees from the setting field direction and on average the CoFe has only rotated about 20 degrees. The histogram has been fit using a Boltzman distribution shown by the blue line. The interface mixing was varied from 0.1 nm - 1 nm in steps of 0.1 nm, with five simulation being run at each value, totalling fifty simulations.}
\label{fig:mix_angles}
\end{figure}

The first step is to confirm that the exchange bias still exists when the interface is mixed. The first challenge in running a hysteresis loop with a mixed interface comes from the setting process. The setting process used for the flat interface from Jenkins \etal~\cite{JenkinsGranular2020} calculates the number of atoms in each sublattice at the interface. In a mixed interface system there is no longer only one interface layer and it is no longer a simple calculation to work out the setting direction. Instead it was found that in a system with a mixed interface, following an experimental setting procedure, the direction of the CoFe magnetisation remained approximately along the setting field direction. The first step of the simulation is field-cooling, the system is heated to a high temperature (above the \Neel temperature) and cooled to 0K under a 0.1T field. 
To check that the setting wasn't a happy statistical accident, the simulation was repeated for 50 structures with interface mixing widths varying from 0.1nm to 1nm. The resulting angle from the setting field direction after the second step - the equilibration stage is plotted in Fig. \ref{fig:mix_angles}. The figure shows that the maximum rotation from the setting field direction was 35 degrees, but the majority of the simulations remained within 20 degrees of the setting field direction. An angle of 35 degrees means that the magnetisation is still 80\% along the setting field direction. The small rotation from the setting field direction is expected and observed experimentally~\cite{OGrady2010AFilms}.

The interface mixing causes the CoFe and IrMn to have more neighbours of the opposite type, meaning the number of interface exchange interactions is higher. This will naturally increase the coupling between the CoFe and the Mn meaning the field between the CoFe and the IrMn is higher, which could be why the setting procedure works in mixed interface systems but not flat systems. Now that the setting procedure has been proven to work in our mixed interface systems, the exchange bias can be simulated. The first step in our simulation is to calculate the dependence of the exchange bias field and the coercivity on a system with a mixed interface and then from this determine the training.

\begin{figure}[!tb]
\centering
\includegraphics[width=0.5\textwidth]{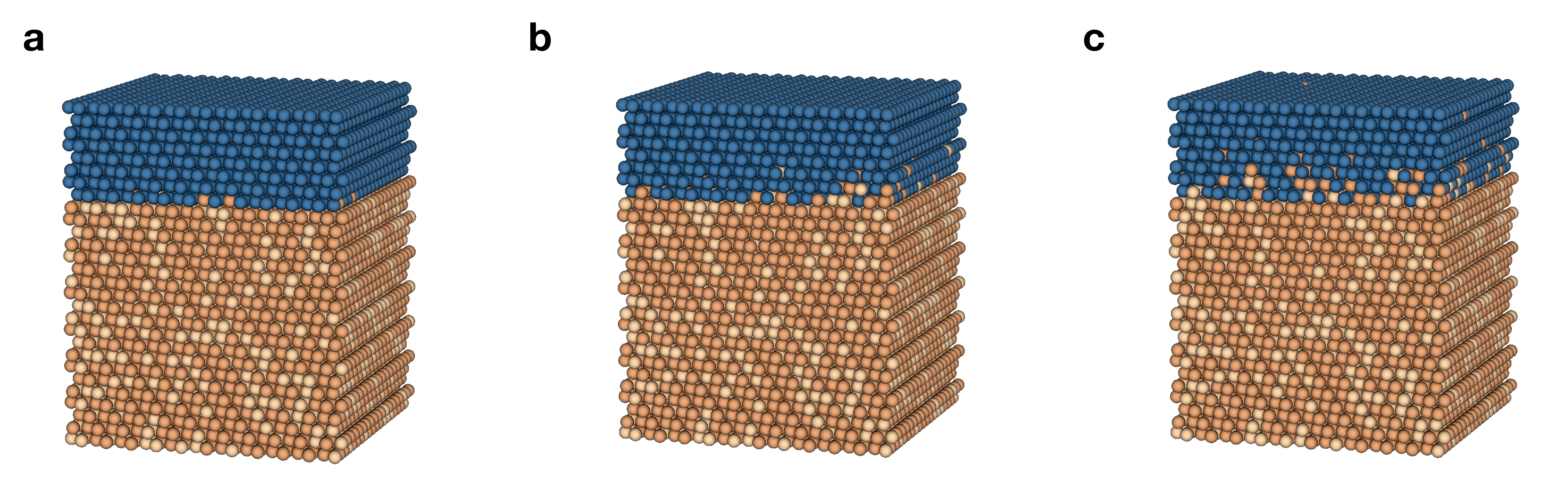}
\caption[Visualisation of different interface mixing widths in an IrMn/CoFe bilayer]{\textbf{Visualisation of different interface mixing widths in an IrMn/CoFe bilayer.} Interface mixing of (a) 0.1nm, (b) 0.5nm and (c) 1nm. }
\label{fig:mixing_width}
\end{figure}

The same multi-granular structure from Jenkins \etal~\cite{JenkinsGranular2020} was used for these simulations so the exchange bias can be compared to the flat interface case. The simulation was a 0K simulation, so only the athermal training effect is accounted for. The width of the interface mixing distribution was systematically varied from 0.1 nm to 1 nm in 0.1 nm intervals totalling 10 different values for interface mixing width. For each value five simulations were run. The five simulations all had exactly the same granular structure, however, the random number seed which defines the specific interface mixing structure was changed. A visualisation of a subsection of one grain of the bilayer in x,y is shown for interface mixing widths of 0.1 nm, 0.5 nm and 1 nm in Fig.~\ref{fig:mixing_width} (a), (b) and (c) respectively. The simulated systems are run through the same simulation steps with the setting field along the $x$ direction. The systems are each cooled from above the \Neel temperature under the presence of an applied field, then the system is left to equilibrate under no field at zero Kelvin, and finally a hysteresis loop is simulated. 

\subsection{Equilibration stage}
\begin{figure}[!tb]
\centering
\includegraphics[width=0.5\textwidth]{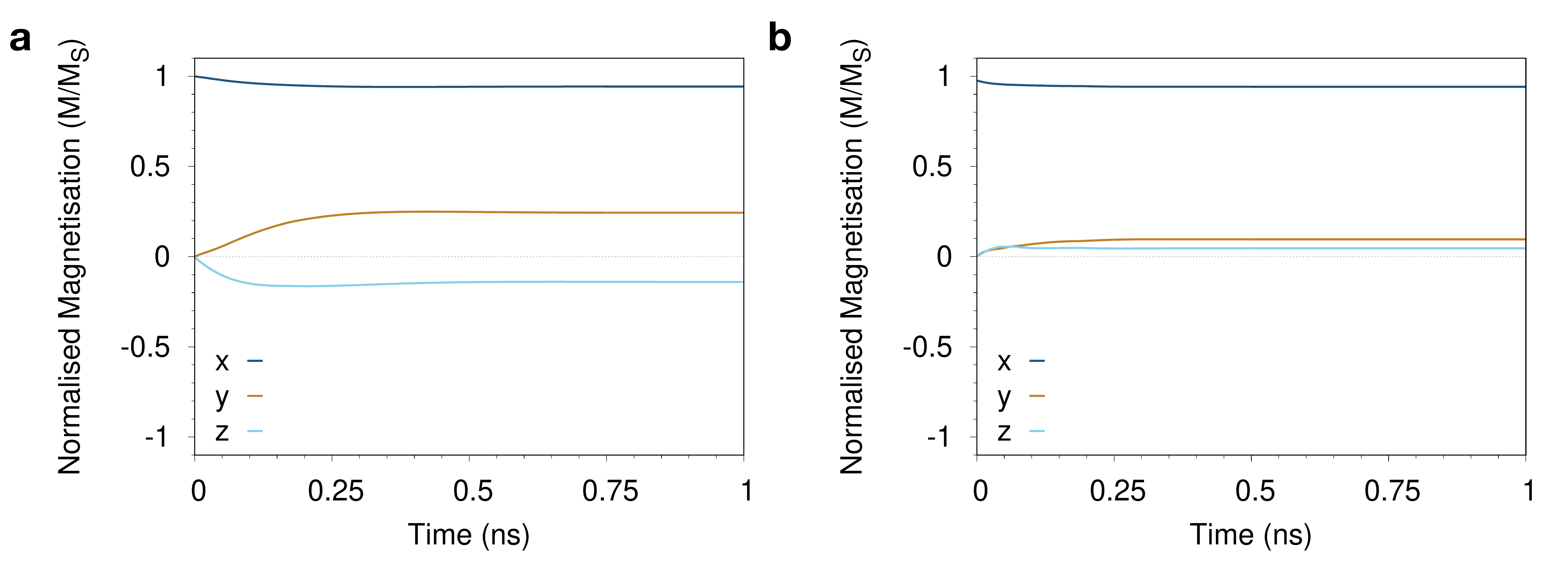}
\includegraphics[width=0.5\textwidth]{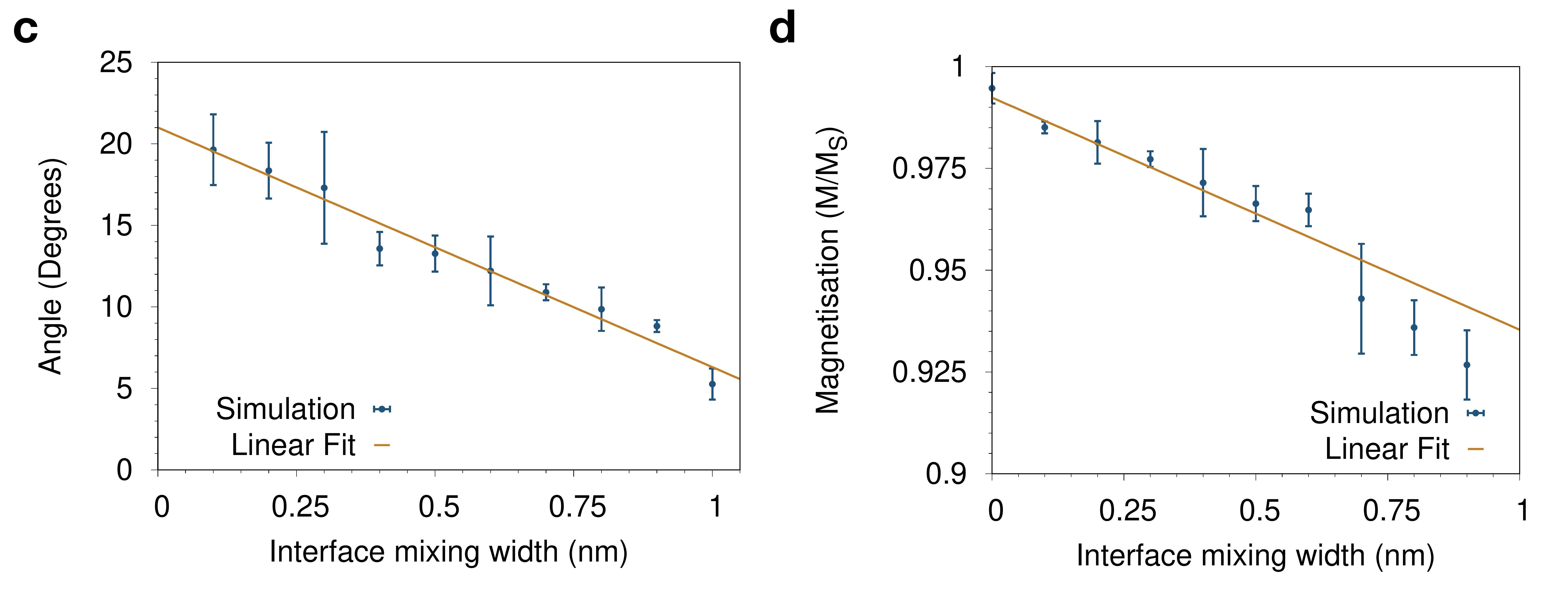}
\caption[Magnetisation vs time data, setting angle and order parameter for the CoFe layer during the equilibration simulation]{\textbf{Magnetisation vs time data for the CoFe layer during the equilibration simulation.} For interface mixing widths of (a) 0.1nm, (b) 1nm. (c) The angle between the CoFe magnetisation at the end of the equilibration simulation and the setting field direction. There is a linear fit to guide the eye. (d) The magnetisation length at the end of the equilibration simulation. }
\label{fig:mix_eq}
\end{figure}

A plot of magnetisation vs time for the CoFe during the equilibration stage is shown in Fig. \ref{fig:mix_eq}, comparing the interface mixing widths of (a) 0.1 nm (b) 1 nm. The simulation with an interface mixing of 0.1 nm has canted to almost 30 degrees away from the setting field direction whereas the 1 nm simulation has remained almost perfectly aligned along the setting field direction. 
 
%\begin{figure}[!tb]
%\centering
%\includegraphics[width=0.5\textwidth]{mag_eq.pdf}
%\caption[The mean angle between the magnetisation at the end of the equilibration and the setting field direction of the CoFe after the equilibration stage and the length of the magnetisation.]{\textbf{The mean angle between the magnetisation at the end of the equilibration and the setting field direction of the CoFe after the equilibration stage and the length of the magnetisation.} (a) The angle between the CoFe magnetisation at the end of the equilibration simulation and the setting field direction. There is a linear fit to guide the eye. (b) The magnetisation length at the end of the equilibration simulation.}
%\label{fig:eq_mix_mlmx}
%\end{figure}

The average magnetisation directions for the CoFe at the end of the equilibration simulation was calculated for each of the 50 simulations and the trend is shown in Fig.~\ref{fig:mix_eq} (c). The figure shows that as the interface mixing increases the angle from the setting field decreases and the system is more strongly set along the setting field direction. It can also be observed that as the interface mixing is increased the magnetisation length of the CoFe decreases as shown in \ref{fig:mix_eq} (d). The decrease in magnetisation length suggests that the CoFe spin directions become disordered at the interface. 

\begin{figure}[!tb]
\centering
\includegraphics[width=0.5 \textwidth]{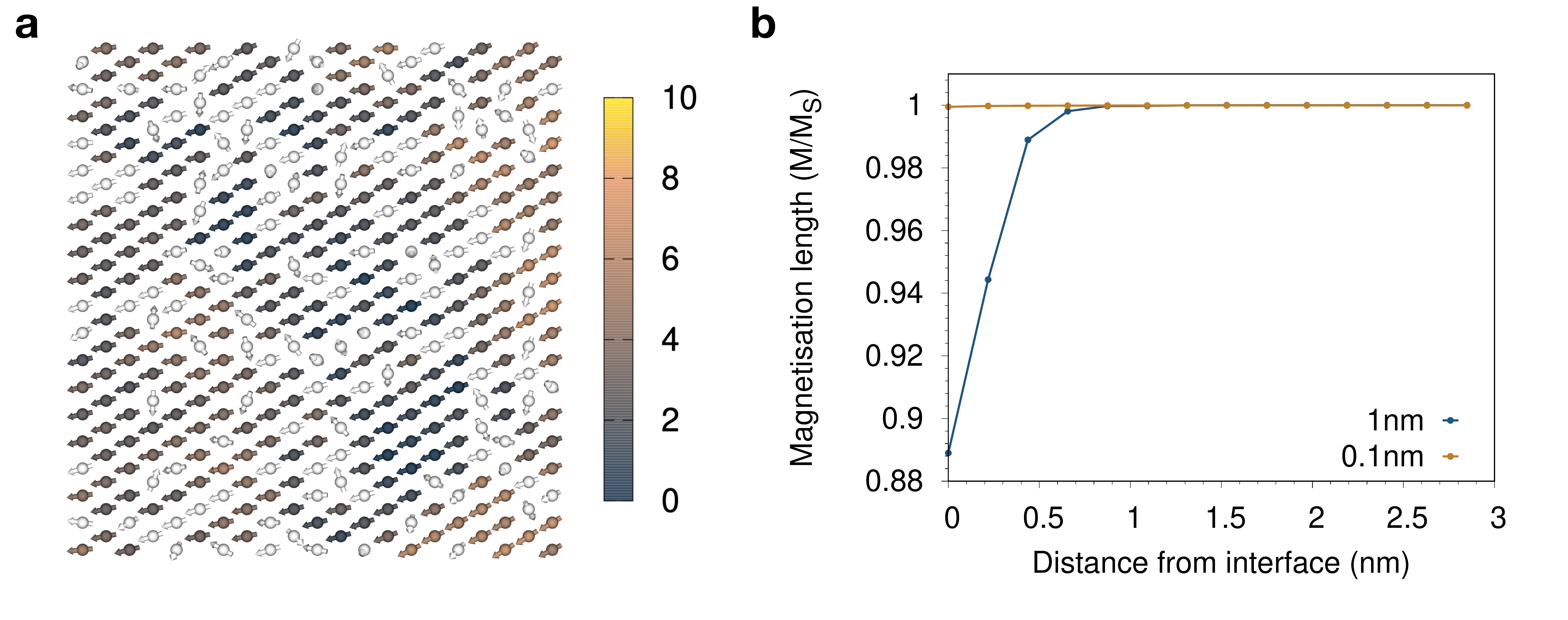}
\caption{\textbf{The interface structure of the CoFe for an interface mixing of 1nm and the magnetisation length for each CoFe layer for interface mixing widths of 0.1nm and 1nm} (a) The magnetisation direction of the CoFe at the end of the equilibration simulation for a 8 nm $\times$ 8nm section of the bilayer. The white arrows represent the Mn spins and the coloured atoms represent the CoFe spins. The colour of the CoFe spin correspond to the angle in degrees from the average direction of the CoFe. Some of the CoFe spins have rotated up to about 10 degrees from the average field direction. (b) The magnetisation length of each CoFe layer, for the simulation with 0.1 nm of interface mixing the CoFe is perfectly aligned at each atomic layer. For the simulation with 1 nm of interface mixing the CoFe is disordered for about 1 nm, then is completely ordered, the magnetisation length of the interface layer is approximately 89\%.}
\label{fig:eq_mix_mlmx_int}
\end{figure}

To investigate the cause of the decrease in magnetisation length the interface spin structure for the simulation with an interface mixing width of 1 nm is shown in Fig.~\ref{fig:eq_mix_mlmx_int} (a). The interface spin structure for the CoFe is no longer completely magnetised along the same direction. Instead at the interface the CoFe has canted up to 10 degrees away from the average magnetisation direction of the CoFe. The canting can be seen to be more prevalent in areas where there are more Mn atoms nearby. In these areas the CoFe is less coupled to the bulk CoFe and is instead coupled to the Mn causing the CoFe to cant towards to Mn spin directions. The magnitude of this disorder was measured by summing the magnetisation of the CoFe atoms in each atomic layer. From this the magnetisation length is calculated as shown in Fig.~\ref{fig:eq_mix_mlmx_int} (b). For an interface mixing width of 1 nm, far from the interface the CoFe has a magnetisation length of one, but near the interface the magnetisation length has decreased to only 89\%. The decrease in magnetisation is most prominent in the interface layer and only occurs for atomic planes up to  1 nm After 1 nm there will only be a small amount of mixing between the CoFe and the Mn and every CoFe atom will be strongly coupled to the bulk CoFe. For an interface mixing of 0.1 nm, the CoFe is completely magnetised at all atomic planes as in all layers the CoFe atoms can couple to the bulk CoFe. Thus, we can conclude that the increased interfacial mixing leads to spin frustration in the CoFe layer, as well as a highly disordered spin state in the interfacial Mn spins.

\subsection{Simulations of the first hysteresis loop}

\begin{figure}[!tb]
\centering
\includegraphics[width=0.48\textwidth]{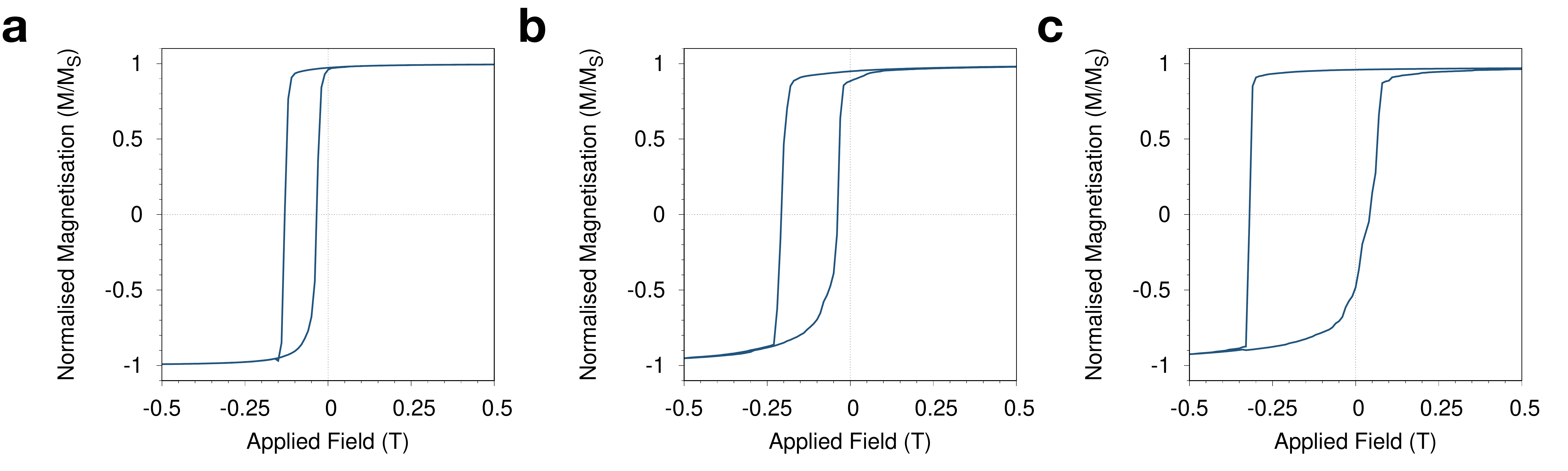}
\caption[First hysteresis loops for multigranular simulations with interface mixing]{\textbf{First hysteresis loops for multigranular simulations with interface mixing}. Hysteresis loops for interface mixing widths of (a) 0.1 nm, (b) 0.5 nm and (c) 1 nm.}
\label{fig:mixhy_mg1}
\end{figure}

The hysteresis loop simulations were run along the magnetisation direction of the CoFe after the equilibration simulation to give the maximum possible exchange bias in each system. The simulations were again run at zero Kelvin to remove any thermal training effects. Simulated hysteresis loops for interface mixing widths of 0.1 nm, 0.5 nm and 1 nm are shown in Fig.~\ref{fig:mixhy_mg1}. The most noticeable difference between the three hysteresis loops is the massive increase in coercivity as the interface mixing width becomes larger. The coercivity has increased from 0.082 T for the 0.1 nm simulation to 0.32 T for the 1 nm simulation. The exchange bias has also increased between the three simulations from 0.09 T for 0.1 nm to 0.15 T for the 1 nm simulations.

\begin{figure}[!tb]
\centering
\includegraphics[width=0.48\textwidth]{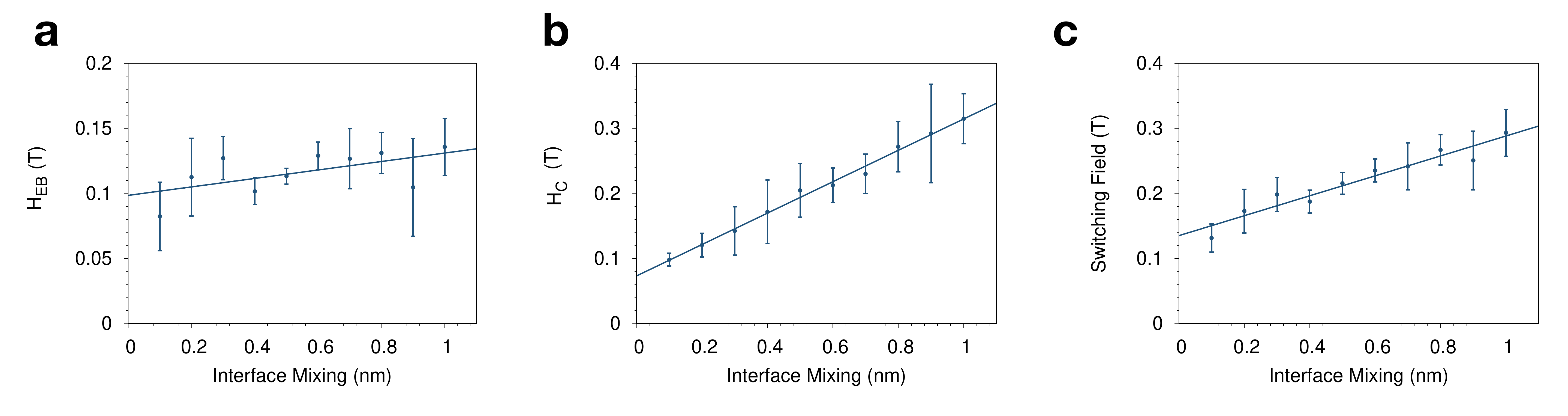}
\caption[The dependence of the exchange bias and coercivity on the interface mixing for the first simulated hysteresis loops]{\textbf{The dependence of the exchange bias and coercivity on the interface mixing for the first simulated hysteresis loops} (a) Shows the dependence of the exchange bias on the interface mixing. (b) Shows the dependence of the coercivity on the interface mixing. (c) The dependence of the switching field on the interface mixing. In all of the figures, the error is the standard deviation in the points from five simulations. All of the figures have been fit with straight lines to help guide the eye.}
\label{fig:mix_ebhc1}
\end{figure}

For each level of interface mixing we performed 5 simulations with different atomic structures, computing the hysteresis loop to determine the exchange bias and then analysed the data to calculate the mean values and standard deviation for the exchange bias, coercivity and switching field. The average exchange bias for each interface mixing width is plotted in Fig.~\ref{fig:mix_ebhc1}(a) showing that there is a large range of exchange bias values, and that the exchange bias can be seen to slightly increase with interface mixing, but the standard deviation sizes means that there is not much of an upward trend. The coercivity is plotted in Fig.~\ref{fig:mix_ebhc1}(b) showing a massive increase to almost 0.3 T for simulations with a high level of interface mixing. The increase in coercivity is due to the fact that with interface mixing a larger proportion of the bulk Mn is incorporated into the interface. Most of these Mn spins are strongly coupled to the bulk IrMn and so have a very large anisotropy, but as they are also strongly exchange coupled they from a reversible interfacial moment~\cite{JenkinsEB2020}, thus causing an increase in the interfacial anisotropy. Fig.~\ref{fig:mix_ebhc1}(c) shows the change in the switching field. The switching field represents the stability of the CoFe to an applied field in the opposite direction. The first switching field (H$_{C1}$) has increased with an increased interface mixing width. There is a noticeable correlation between the coercivity and the first switching field value, suggesting that the second switching field (H$_{C2}$) is not affected by the interface mixing. The interface mixing instead only affects the first switching field.

The experimental dependence of exchange bias on interface roughness is still not quantified~\cite{Qi2019InfluenceMultilayers}. There have been many experimental measurements but as IrMn is naturally disordered and the interface mixing will have an effect on the interface spin configurations it is hard to quantify. Qi \etal~\cite{Qi2019InfluenceMultilayers} experimentally measured that there was no correlation between interface mixing and exchange bias value, for interface roughness of 0.678 nm, 0.823 nm and 1.259 nm whereas, Parkala \etal~\cite{Choo2007ABilayers} measured a decrease in exchange bias with an increase in interfacial roughness comparing interface roughness values of 0.1 nm and 1.1 nm. Our simulations do not agree with either of these results, however, both of these measurements were done at non-zero temperature and therefore this may cause a large decrease in the exchange bias in the structures with more disorder due to thermal spin fluctuations.

\subsection{Simulations of the second and third hysteresis loops}
\begin{figure}[!tb]
\centering
\includegraphics[width=0.5\textwidth]{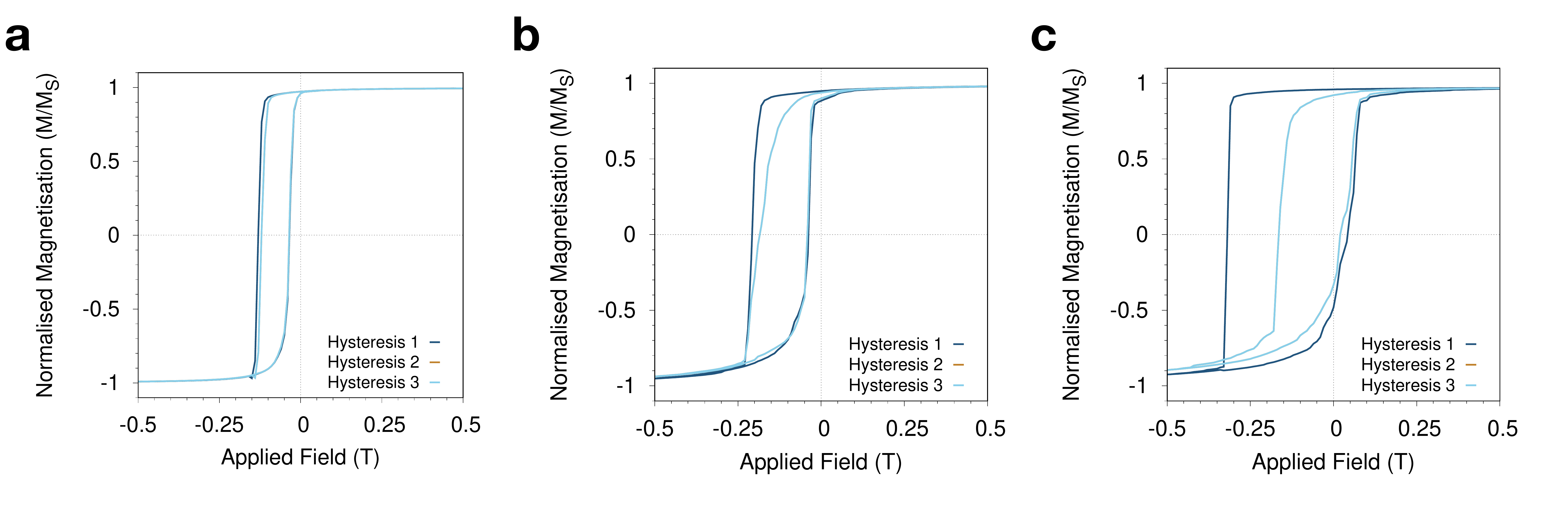}
\caption[The first three simulated hysteresis loops for different intermixing widths]{\textbf{The first three simulated hysteresis loops for different intermixing widths} The interface mixing widths were (a) 0.1nm, (b) 0.5nm and (c) 1nm.}
\label{fig:traininghy2}
\end{figure}

Now that we have proven that exchange bias still exists in systems with mixed interfaces we can see if the mixing has caused the hysteresis loops to exhibit the training effect. To investigate this two more hysteresis loops have been run on the simulated system, both at zero Kelvin. These hysteresis loops are shown in Fig. \ref{fig:traininghy2}, the first, second and third simulated hysteresis loops are shown for interface mixing widths of 0.1 nm, 0.5 nm and 1 nm. All of the hysteresis loops exhibit a decrease in the exchange bias and the coercivity between the first and second measured hysteresis loops, analogous with the training effect. The magnitude of the decrease in both the coercivity and the exchange bias is observed to increase with the width of interface mixing. The change in the exchange bias between the first and second simulated hysteresis loops for an interface mixing width of 0.1 nm is almost negligible. Whereas, the exchange bias in the 1 nm simulation has decreased dramatically between the first and second simulated hysteresis loops. In all three systems there is approximately no change in either the coercivity or the exchange bias between the second and third simulated hysteresis loops and no change in the coercivity. This agrees with previous experimental measurements of low temperature systems, where a large decrease in the exchange bias is found between the first and second measured hysteresis loops only~\cite{Kaeswurm2011TheFilms}. 

\begin{figure}[!tb]
\centering
\includegraphics[width=0.5\textwidth]{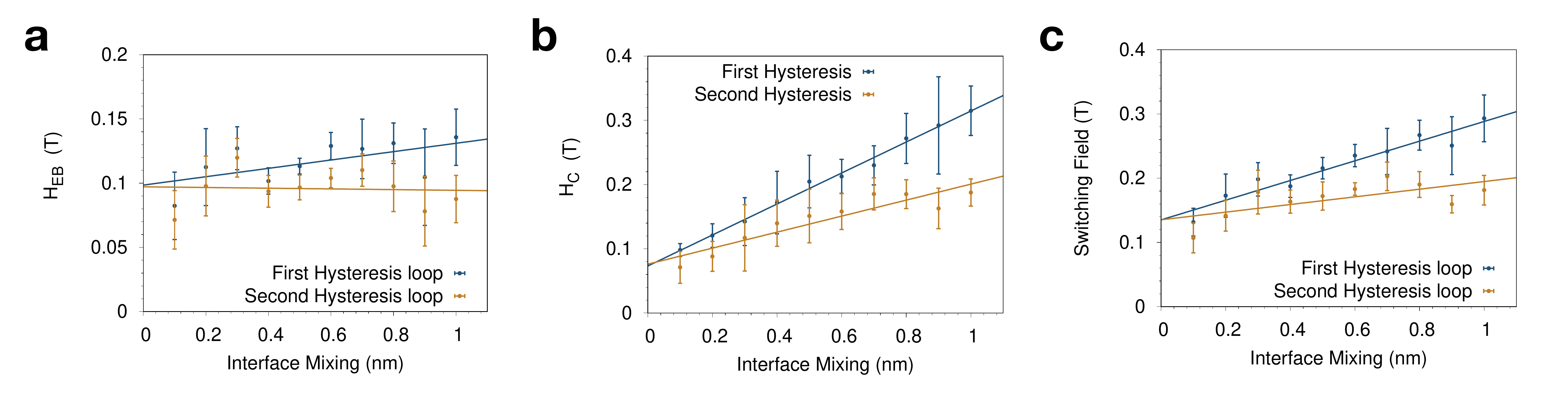}
\caption{\textbf{The dependence of the exchange bias and coercivity on the interface mixing for the first and second simulated hysteresis loops} (a) Shows the dependence of the exchange bias on the interface mixing. (b) Shows the dependence of the coercivity on the interface mixing. (c) The dependence of the switching field on the interface mixing. For all of the figures the first and second hysteresis loops are shown and the error is the standard deviation in the points. All of the figures have been fit with straight lines to help guide the eye.}
\label{fig:mix_ebhc2}
\end{figure}

The second and third hysteresis loops were simulated for all of the fifty systems. Fig.~\ref{fig:mix_ebhc2} shows that for all interface mixing widths the exchange bias and the coercivity has decreased between the first and second simulated hysteresis loop. The most noticeable change is that for the second hysteresis loops the larger the interface mixing the larger the decrease in exchange bias between the first and second loops. For the second hysteresis loop all of the simulations gave very similar values for the exchange bias field, meaning that there is no correlation between exchange bias and interface mixing width, as experimentally predicted by Qi \etal~\cite{Qi2019InfluenceMultilayers}. For the coercivity, the higher the interface mixing the higher the drop in coercivity between the first and second simulated hysteresis loops. This leads to a plateau in the coercivity values at about 0.18 T for high levels of interface mixing. Fig. \ref{fig:mix_ebhc2}(c) shows that the first switching field 
increases with increased interface mixing, but plateaus at about 0.18T again showing a similar form to the coercivity. The initial increase of exchange bias with interface mixing increases the stability of the CoFe. However, this plateaus because only Mn atoms which are still coupled to the bulk Mn will increase the coercivity, once the interface mixing it too high the Mn atoms are no longer coupled to the bulk and therefore don't contribute as there is no difference between the second and third simulated hysteresis loops.

\begin{figure}[!tb]
\centering
\includegraphics[width=0.5\textwidth]{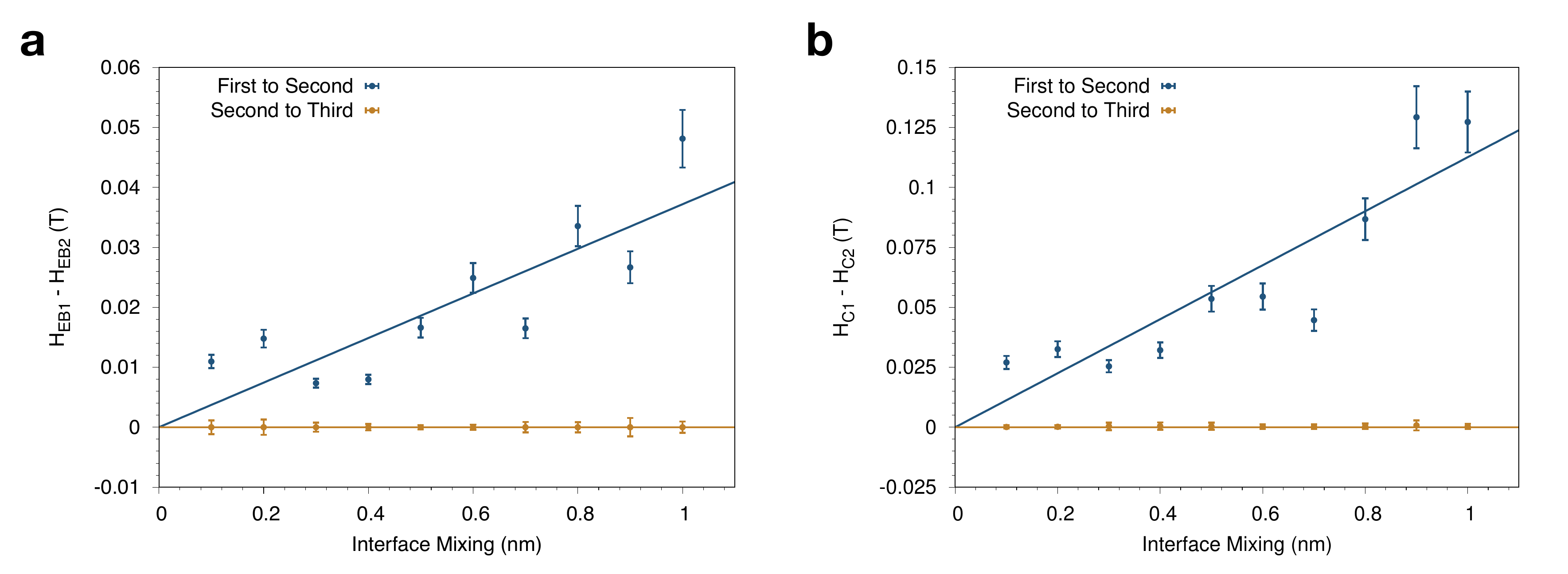}
\caption[The change in exchange bias and coercivity between the first and second and second and third simulated hysteresis loops]{\textbf{The change in exchange bias and coercivity between the first and second and second and third simulated hysteresis loops} (a) The change in the exchange bias between consecutive hysteresis loops, with lines of best fit to guide the eye. (b) The change in the coercivity between consecutive hysteresis loops, with lines of best fit to guide the eye. There is a large change in both the coercivity and the exchange bias between the first and second simulated hysteresis loops but almost no change between the second and third. }
\label{fig:mixEBHC}
\end{figure}

The mean change in exchange bias and coercivity between the first and second and second and third hysteresis loops is shown in Fig.~\ref{fig:mixEBHC}. The error is the standard deviation in the values. The simulations have shown there is a large decrease in both the exchange bias and the coercivity between the first and second hysteresis loops, but almost no change between the second and third simulated hysteresis loops. The change in both the coercivity and the exchange bias is proportional to the width of the interface mixing. The more mixed the interface the higher the change between the first and second hysteresis loops. Experimentally, a continuous decrease is measured due to the thermal training effect. From our simulations we have demonstrated that the athermal training effect only occurs between the first and second measured hysteresis loops, and is therefore truly athermal. 

\subsection{The interface structure throughout the hysteresis loops} 
\begin{figure}[!tb]
\centering
\includegraphics[width=0.5\textwidth]{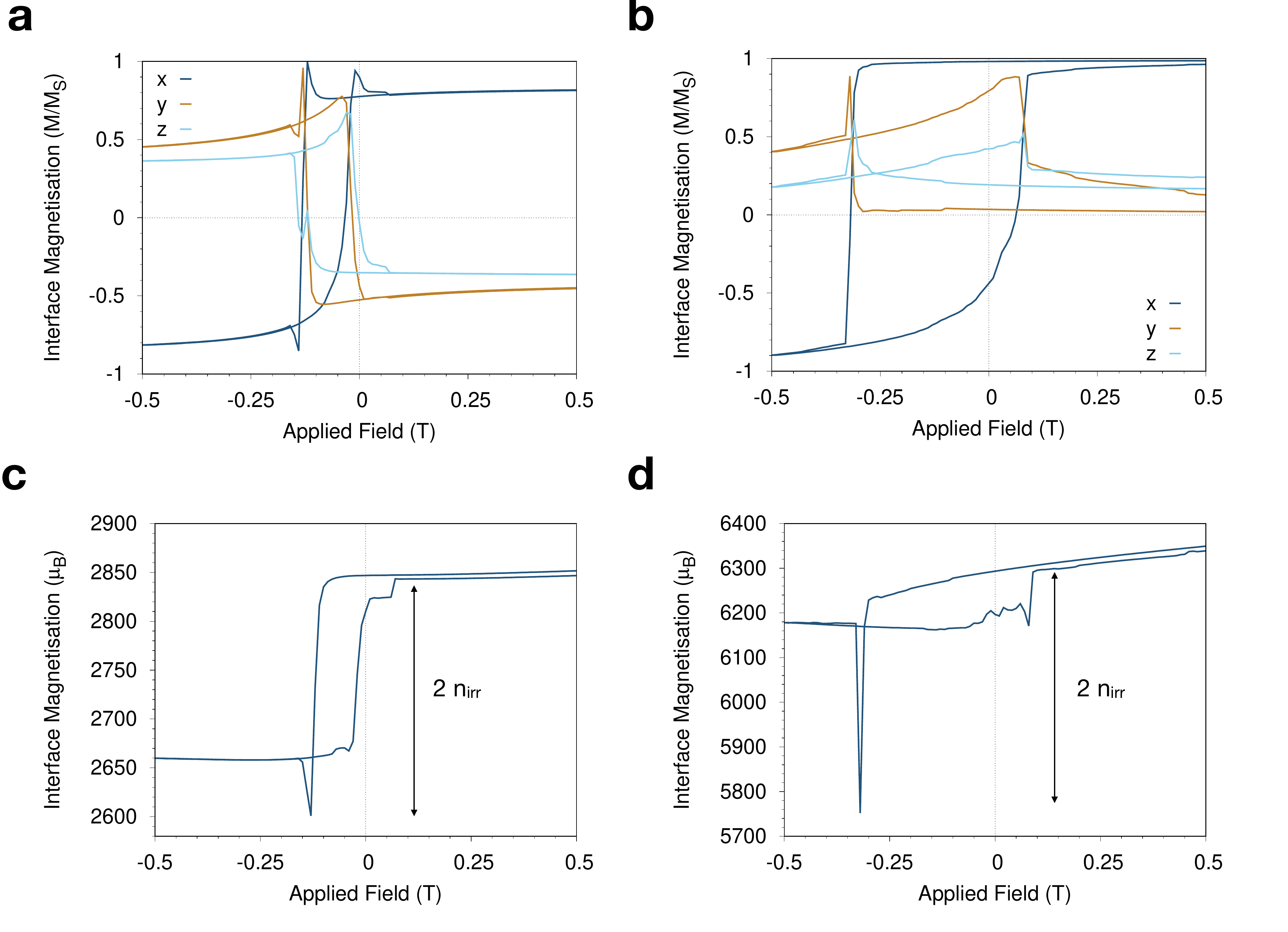}
\caption[The magnitude and direction of the interface moment of the Mn throughout the hysteresis loop]{\textbf{The magnitude and direction of the interface moment of the Mn throughout the hysteresis loop.} The direction of the Mn moment in the interface layer (in direct contact with the CoFe) throughout the hysteresis loops for interface mixing widths of (a) 0.1 nm and (b) 1 nm respectively. In both cases the interface moment has followed the interface moment of the CoFe (shown in Fig. \ref{fig:mixEBHC}). (c) and (d) The magnitude of the net interface moment throughout the hysteresis loops for interface mixing widths of 0.1 nm and 1 nm respectively.}
\label{fig:mix_ms_int}
\end{figure}

To understand what is causing the training effect, the magnetisation in the interface layer was observed throughout the hysteresis loop for the simulated systems with intermixing of 0.1 nm and 1 nm shown in Fig.~\ref{fig:mixhy_mg1} and Fig.~\ref{fig:traininghy2}. The direction and magnitude of the interface magnetisation of the Mn is shown in Fig.~\ref{fig:mix_ms_int}, where the direction of the magnetisation can be seen to follow the direction of the CoFe magnetisation as they are ferromagnetically coupled together. The interface magnetisation decreases between positive and negative saturation as was observed by Jenkins \etal~\cite{JenkinsEB2020}. The decrease in saturation magnetisation of the CoFe at the zero-field point is due to the number of irreversible Mn spins ($n_{irr}$) in the interface layer.

Both interfaces have a pronounced minima in the interface moment just after the first switch has occurred, suggesting a large reordering of the interface spins at this point. This reordering is analogous to the meta stable spins described by Biternas \textit{et al}~\cite{Biternas2009StudyModel}. At the start of the first hysteresis loop the interface magnetisation of the Mn is in a meta-stable state, which arose during the setting procedure. It takes a large field to evolve the interface magnetisation from this meta stable state into the ground state - causing the pronounced minima in the interface moment. As the change in spin structure occurs just after the first switching point of the hysteresis loop, by the time the system has reached the negative saturation point it is already in the ground state configuration and no longer in a meta stable state. This explains why the returning loops of the first and second hysteresis loops are always experimentally observed to have a similar shapes whereas there is a large change in the first arm of the hysteresis loops~\cite{Biternas2009StudyModel}. 

For the 0.1 nm interface, the pronounced minimum in the interface magnetisation (2603$\mu_B$) is about 20\% lower than the negative saturation value for the interface magnetisation (2662$\mu_B$). We therefore expect the exchange bias to be about 20\% lower in the second hysteresis loop simulation as it no longer has to overcome this larger energy barrier. For the 1 nm simulation the pronounced minima (5765$\mu_B$) is about 70\% lower than the interface magnetisation at negative saturation (6183$\mu_B$). It is therefore expected that the exchange bias will decrease by about 70\% between the first and second hysteresis loops. Both of these predictions approximately match the change in exchange bias shown in the hysteresis loops shown in Fig.~\ref{fig:mixhy_mg1}. 

\begin{figure}[!tb]
\centering
\includegraphics[width=0.5\textwidth]{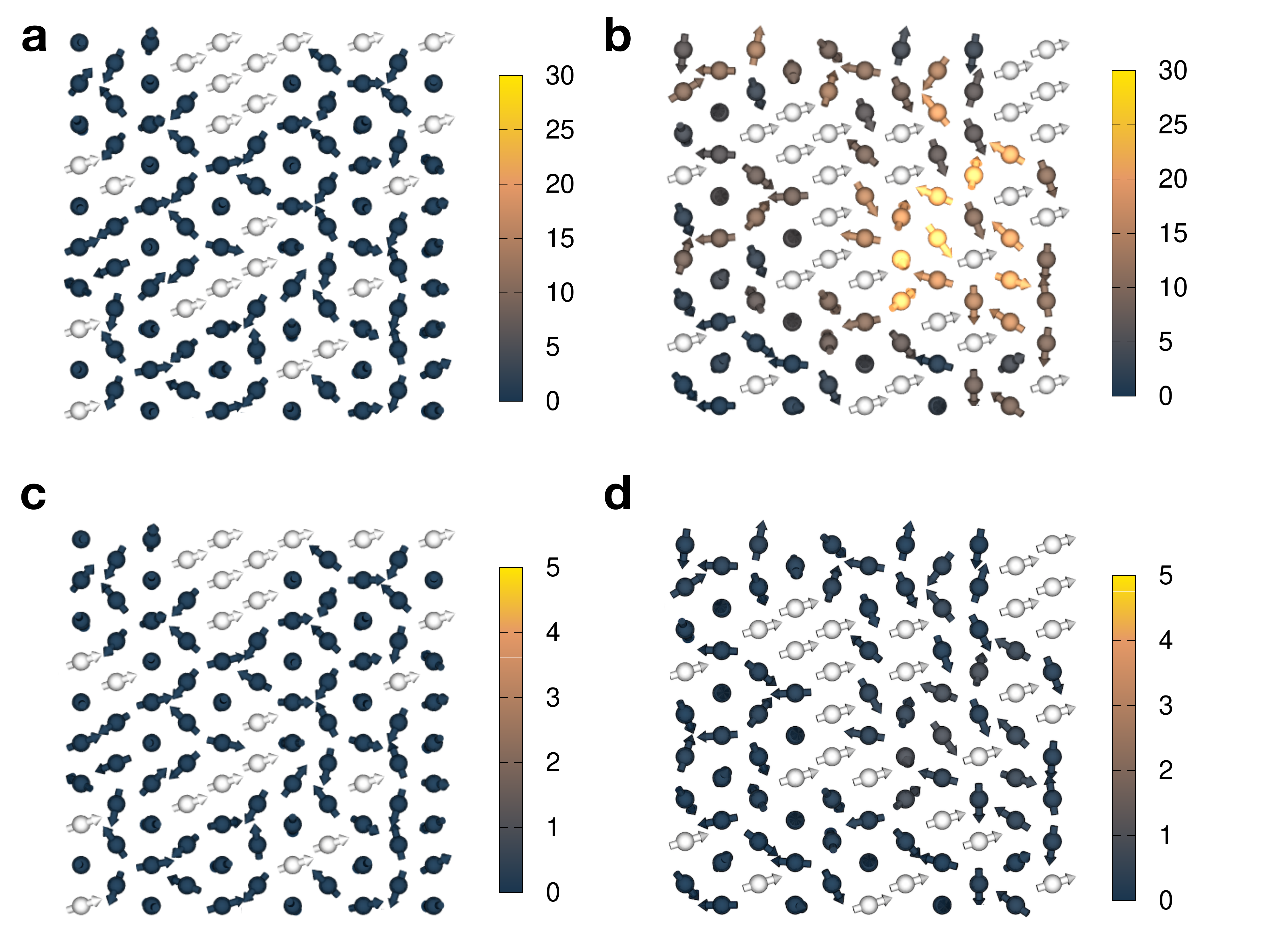}
\caption{\textbf{The change in the interface spin structures between the start and the end of the hysteresis loops.} (a) and (b) show the change in interface spin structure between the start and end of the first hysteresis for interface mixing widths of 0.1 nm and 1 nm respectively. (c) and (d) show the change in interface spin structure between the start and end of the second hysteresis for interface mixing widths of 0.1nm and 1nm respectively. The change in colour shows the change in angle in degrees as shown by the scale bar on the side.}
\label{fig:move}
\end{figure}

In Fig. \ref{fig:mix_ms_int} we observe that in both the 0.1 nm and 1 nm values of interface mixing there is a change in  magnitude of the Mn interface magnetisation between the start and the end of the first hysteresis loop. In Fig.~\ref{fig:mix_ms_int}(b) the direction of the interface moment has also changed from (a). This is because the interface configuration has changed from a meta-stable state to the ground state. The change in spin configuration means that the first and second hysteresis loops will start from different interface spin structures. To quantify this change, a subsection of the interface spin structure was visualised and the change in angle from start to end of the hysteresis loop was calculated. The angles between the initial and final positions of the spins are shown in Fig.~\ref{fig:move}. From this image it can be seen that for low levels of interface mixing there is almost no rotation of Mn spins between the initial and final states of the hysteresis loop. However, for the larger interface mixing widths it can be seen that there has been a large level of distortion between the initial and final states. The change in the interface spin structure means that the interface will have a different number of reversible and irreversible spins compared to the initial hysteresis loop. After this the interface has reordered. Fig.~\ref{fig:move} also shows that there is no large change in angle for either the 1 nm or 0.1 nm simulated system between the start and end of the second hysteresis loop. This shows that the spin configuration has returned to the ground state and the spin configuration has become stable.

\begin{figure}[!tb]
\centering
\includegraphics[width=0.5\textwidth]{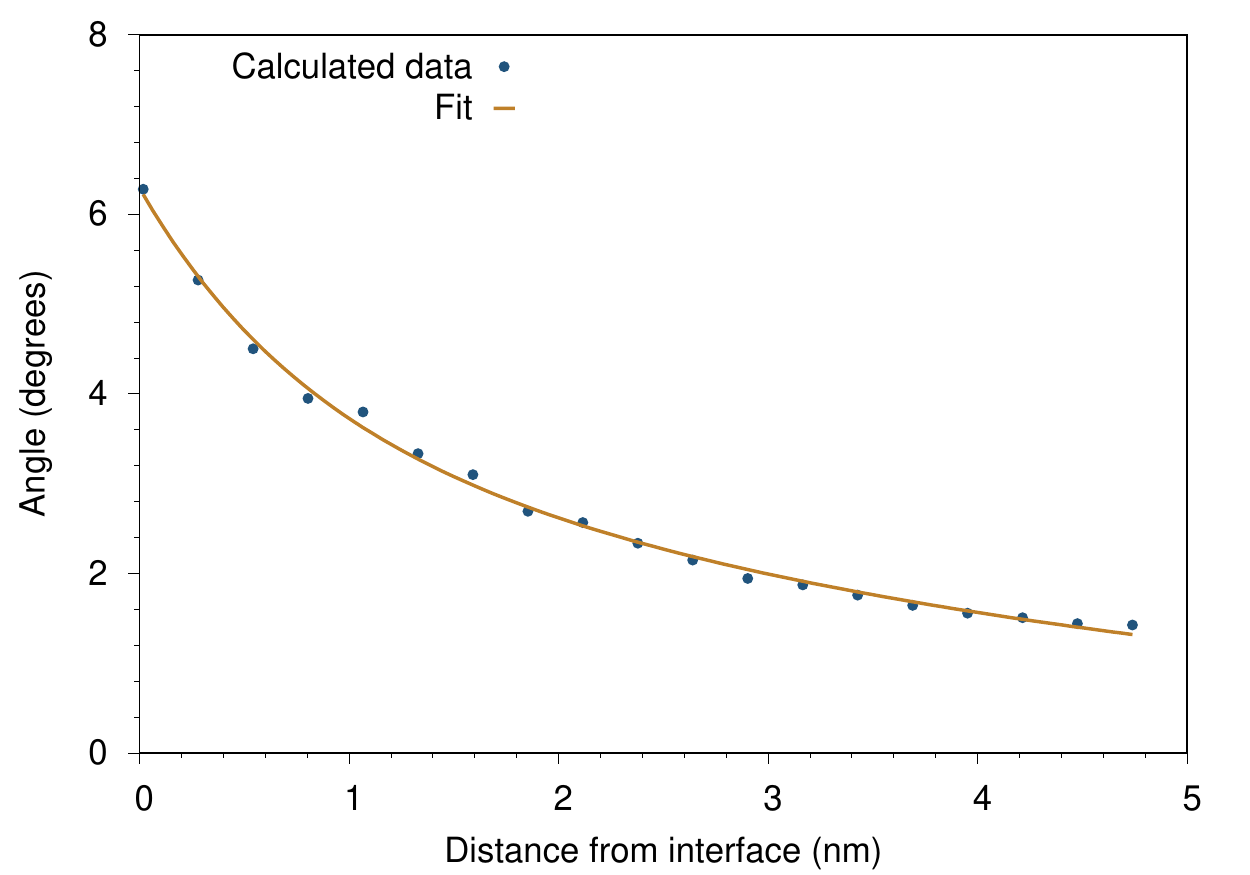}
\caption[The average rotation of each layer of the Mn between the start and end of the first hysteresis loop]{\textbf{The average rotation of each layer of the Mn between the start and end of the first hysteresis loop.} The angle between the start and end magnetisation of each Mn spin during the first hysteresis loop. These were averaged across every Mn layer. The interface spins have a much higher change in magnetisation that the bulk Mn.}
\label{fig:angles}
\end{figure}

To work out whether this reordering is due to a movement of the entire bulk structure or just an interface effect the average angle between initial and final hysteresis loop was plotted as a function of distance from the interface. The interface used has an interface mixing width of 1nm, to show the largest changes in angle as it is assumed this will have the largest effect on the bulk Mn spins.  The angles are shown in Fig.~\ref{fig:angles}, the plot shows the average angle the atoms in each layer have moved. The interface spins have moved an average of approximately 6 degrees at the interface but far away from the interface the spins have only rotated about 2 degrees. This suggests that the movement is predominantly an interface effect and not a bulk effect. This is to be expected due to the large anisotropy in IrMn, and so in this system domain wall effects within the antiferromagnetic layer play a very small role in the reversible interfacial moment.

Finally we consider the qualitative differences in the spin structure that give rise to the athermal training effect. The exchange bias in IrMn/CoFe systems is always determined by the combination of reversible and irreversible interfacial spin moments~\cite{JenkinsEB2020,JenkinsGranular2020}. In the case of intermixing, there is a much greater degree of coupling between the CoFe and IrMn, with Mn spins embedded in the CoFe still coupled strongly to the bulk IrMn giving them a large degree of metastability, and thus a very large initial exchange bias. However, in the first hysteresis cycle, when forcing the CoFe spins to rotate, the high degree of exchange coupling overcomes the anisotropy of the Mn spins, forcing a reordering of the spin structure as seen in Fig.~\ref{fig:move} and the large transient in the size of the interfacial magnetization at the switching field. This reversing process is sufficient to destabilise the irreversible spins and causes them to become reversible and more closely follow the direction of the CoFe layer rather than the IrMn as is the case immediately after the setting procedure. Thus, the athermal training effect arises due to the conversion of initially irreversible spins to reversible spins, enabled by the weaker coupling of Mn spins in the CoFe layer to the underlying IrMn layer.

\section{Conclusion}

The development of novel spintronic devices hinges on the maximisation of the exchange bias effect. A large problem in maximising exchange bias is the athermal training effect, which is a large drop in exchange bias between the first and second measured hysteresis loops. In this paper we have used an atomistic model to determine the cause of the athermal training effect. We have found that the drop in exchange bias is caused by meta stable spin states occurring during the setting process, which are more weakly coupled to the IrMn layer and convert to reversible spins after the first hysteresis cycle. After the first hysteresis loop these spin states are reversed and fall into a minimum energy state, thereby reducing the irreversible interface moment and exchange bias of further hysteresis loops. 

The meta-stable spin states were found to be due to roughness at the interface as no training was found for a perfectly flat interface. The interface can be seen to reorder between the first and second hysteresis loop where the angle between this reordering increases with the amount of interface roughness therefore increasing the training. This shows that training is purely an interface effect, in agreement with the model of Biternas \etal\cite{Biternas2009StudyModel}. So far we have not considered the thermal training effect in this model, as this would require significantly larger simulation volumes and more averaging, but we believe that the standard explanation of the thermal training effect due to the flipping of grains during reversal is essentially correct. This new understanding of the athermal training effect could lead to the development of anti-ferromagnetic spintronic devices with ultra fast dynamics and a robustness to external fields not seen in conventional ferromagnetic devices.

\bibliography{references,library}
\end{document}